\documentclass[aps,prstab,preprint,showpacs,groupedaddress]{revtex4}
\usepackage{graphics,epsfig}

\begin{document}
\def\be{\begin{equation}}
\def\ee{\end{equation}}
\def\bc{\begin{center}}
\def\ec{\end{center}}
\def\bea{\begin{eqnarray}}
\def\eea{\end{eqnarray}}
\draft
\title{Effects of aging and links removal on epidemic dynamics in scale-free networks}
\author{K. P. Chan$^{1}$, Dafang Zheng$^{2}$, P. M. Hui$^{1}$}

\address{$^{1}$Department of Physics, The Chinese University of Hong Kong,
Shatin, New Territories, Hong Kong \\
$^{2}$Zhejiang Institute of Modern Physics, Zhejiang University,
Hangzhou, 310027, People's Republic of China}

\begin{abstract}

We study the combined effects of aging and links removal on
epidemic dynamics in the Barab\'{a}si-Albert scale-free networks.
The epidemic is described by a susceptible-infected-refractory
(SIR) model. The aging effect of a node introduced at time $t_{i}$
is described by an aging factor of the form $(t-t_{i})^{-\beta}$
in the probability of being connected to newly added nodes in a
growing network under the preferential attachment scheme based on
popularity of the existing nodes. SIR dynamics is studied in
networks with a fraction $1-p$ of the links removed. Extensive
numerical simulations reveal that there exists a threshold $p_{c}$
such that for $p \geq p_{c}$, epidemic breaks out in the network.
For $p < p_{c}$, only a local spread results. The dependence of
$p_{c}$ on $\beta$ is studied in detail.  The function
$p_{c}(\beta)$ separates the space formed by $\beta$ and $p$ into
regions corresponding to local and global spreads, respectively.

\end{abstract}


\maketitle


\section{Introduction}
It has been discovered in recent years that many real network
systems, while showing different levels of complexity of their
own, possess novel common structural or topological properties.
These networks include, for example, the Internet
\cite{faloutsos}, World-Wide-Web \cite{huberman}, scientific
citations \cite{redner}, cells \cite{jeong}, the web of
collaborations among actors and actresses \cite{albert}, and the
web of human sexual contacts \cite{liljeros}.  Typically, networks
are characterized by nodes and links.  The nodes may represent
websites on the world-wide-web, a paper in scientific journal, or
an actor, etc., depending on the network concerned.  The links
represent the interaction between the nodes.  The interaction may
represent the links from a webpage to another, references cited in
a journal to papers in the literature, collaboration among actors
and actresses, etc., again depending on the system concerned.
These structures are found to show the small-world effect, i.e., a
node can reach any randomly chosen node through only a few links;
the high-clustering effect, i.e., nodes connected to a chosen node
have high probability of having links among themselves; and a
well-defined degree distribution \cite{albert1,dorogovtsev}. These
universal features observed in real systems have led to the
intensive studies of complex networks in recent years. Physicists
have made major contributions to the understanding of the physics
of networks.  On the structural properties, physicists have
proposed a number of models in which the observed statistical
features in real networks are reproduced. The most representative
of these models are the small-world model proposed by Watts and
Strogatz \cite{watts}; and the scale-free growing network proposed
by Barab\'{a}si and Albert (BA) \cite{barabasi}. The latter model
gives a better description of the scale-free or broad-scale
distributions of degrees found in many real-life networks and
hence has been extensively studied recently.  While structural
properties are important in the geometrical description of
networks, many applications of the physics of networks rely on our
understanding of dynamical processes on networks.  Dynamical
processes, such as percolation, searchability problems, and
especially the spread of diseases on complex networks have been
investigated only recently. It has been found that the BA networks
are highly susceptible to large scale epidemics \cite{pastor}.

In the present work, we study how aging phenomena
\cite{dorogovtsev1} and links removal combined may affect the
spreading of a disease or rumor in BA growing networks. Aging
refers to situations in which the older nodes in a network become
increasingly unattractive to the newly added nodes.  We use a
susceptible-infected-refractory (SIR) model for the dynamics of
epidemics. Using extensive numerical simulations, we study the SIR
dynamics on BA networks with a fraction of $1-p$ of the links
removed or turned ineffective.  It is found that a threshold
$p_{c}(\beta)$ exists, where $\beta$ is an aging parameter. For $p
\geq p_{c}$, a global spread results; while for $p < p_{c}$, only
a local spread results.  For the full BA network, i.e., $p=1$,
global outbreak can take place only if the value of $\beta$ is
smaller than a critical value.

\section{Model}
The underlying network can be set up as follows. Initially there
are $m_{0}$ fully connected nodes. We take $m_{0} = 5$ in the
present work. At each time step, a new node is introduced.  Each
newly added node establishes $m$ ($\leq m_{0}$) outgoing links to
existing nodes. Each new link introduced at time step $t$ has a
probability
\begin{equation}
\label{Probkt}
\Pi_{i}=\frac{k_{i}/(t-t_{i})^{\beta}}{\sum_{l}k_{l}/(t-t_{l})^{\beta}}
\end{equation}
to be connected to an existing node $i$ with degree $k_{i}$, i.e.,
having $k_{i}$ links.  Here $t_{i}$ is the time at which node $i$
was added to the network and $\beta$ is an aging parameter
characterizing how rapidly an existing node becomes unattractive
to a newly added node.  Eq.(\ref{Probkt}) states that newly
established links are preferentially attached to the younger and
highly connected existing nodes in the system. For $\beta = 0$,
there is no aging effect and the model reduces to the BA model
\cite{barabasi}.  The aging factor thus models the situation where
older persons in a society tend to isolate themselves or being
isolated by the younger ones and lose their influence in the
younger persons.

As an undiluted BA network is highly susceptible to large scale
global spread of diseases, it is interesting to study the effects
of removing a fraction of links in a fully grown BA network. These
diluted networks can be constructed as follows. Having set up an
aging network of $N$ nodes, each link has a probability $1-p$ of
being removed. Hence, on the average, a fraction $p$ of the links
in a fully grown aging network consisting of $N$ nodes are kept.
The removed links in a diluted network can be regarded as
ineffective links through which a disease or a rumor cannot
spread.  In the present work, we study aging networks of size up
to $N=50,000$ nodes.

To study epidemiological processes on aging and diluted networks,
we use the three-state Susceptible-Infected-Refractory (SIR) model
\cite{diekmann}.  The model is a standard model for studying
epidemics and the spread of a rumor in a connected population.
Initially, all nodes are in the S(susceptible)-state, and one node
is randomly chosen to be infected, i.e., in the I-state.  The SIR
dynamics proceeds as follows. At a time step $t$, a node, say $i$,
is randomly chosen among all the infected (I) nodes.  A neighbor
or friend $j$ is then selected randomly among all the neighbors of
node $i$, i.e., those with a link connected to $i$. If node $j$ is
susceptible, it becomes infected and the chosen node $i$ remains
in the I-state; otherwise (i.e., node $j$ is either I or R) the
state of node $j$ remains unchanged and the chosen node $i$
becomes refractory (R) at the end of the time step. Within the
context of a spread of a rumor, the I-state nodes refer to persons
who want to spread the rumor, the S-state nodes are persons who
have not heard the rumor, and the R-state nodes refer to persons
who lost interest in spreading the rumor after knowing it. As time
evolves, the number of R-nodes (S-nodes) increases (decreases);
while the number of I-nodes increases initially and then
eventually drops to zero. The number of R-nodes at the end of the
dynamical evolution is denoted by $N_{R}$.  We are interested in
how $N_{R}/N$ varies as the fraction of effective links $p$ and
the aging parameter $\beta$.

\section{Results and Discussion}
We performed extensive numerical simulations to explore the
combined effects of aging and links removal on the SIR dynamics.
In all the simulations, each data point is obtained by averaging
over 100 realizations of the network structure and 100 different
initially infected nodes for each realization. In Fig.1 we show
the mean fraction of R-nodes, $r \equiv <N_{R}>/N$, on networks of
different sizes up to $N=50,000$ on a log-log plot for different
values of the aging parameter (a) $\beta=0$, (b) $\beta=0.5$, (c)
$\beta=1.0$, and (d) $\beta=1.5$. For each value of $\beta$, we
carried out simulations for different levels of dilution $p$. For
global spread, we expect that the number of refractory sites
$N_{R}$ to be directly proportional to the network size $N$ and
hence the fraction of refractory nodes $r$ becomes independent of
$N$.  For local spreads, $N_{R}$ is finite for a sufficiently
large network and hence $r$ decreases as $1/N$. We thus search for
a threshold $p_{c}$ that separate these two behavior of $r$ for a
given value of $\beta$.  The results in Fig. 1 show that $r =
N_{R}/N$ does exhibit different behavior as $p$ decreases. Precise
determination of $p_{c}$ is quite difficult, due to finite size
effect and the inhomogeneous nature of the problem resulting from
the random removal of links from a originally randomly established
BA network.  To proceed, we study the slope of the lines in Fig.1
as $p$ decreases.  Typical results for $\beta = 0$, $0.5$, and
$1.0$ are shown in Fig.2.  Notice that, for given value of
$\beta$, the slope starts to drop from zero, i.e., $N_{R} \sim N$
for global spread, for some value of $p$.  Notice that there is a
transitional range of $p$ for which $N_{R}$ neither scales
linearly with $N$ nor independent of $N$. We adopt the value of
$p$ at which the slope drops to the value of $-0.1$ as
$p_{c}(\beta)$. According to this criteria, we found
$p_{c}(\beta=0)=0.2$, $p_{c}(\beta=0.5)=0.25$,
$p_{c}(\beta=1)=0.3$, and $p_{c}(\beta=1.5)= 0.35$ for the values
of $\beta$ in Fig.1 and Fig.2.

Carrying out the calculations for different values of the aging
parameter $\beta$, we obtained the dependence of the threshold
$p_{c}(\beta)$ on the aging parameter $\beta$.  The results of
$p_{c}(\beta)$ are shown in Fig.3 as the phase boundary in a phase
space formed by the aging parameter $\beta$ and fraction of
effective links $p$. For small $\beta$ ($0 < \beta < 1.25$),
$p_{c}$ increases gradually as $\beta$ increases. For $\beta >
1.25$, $p_{c}$ increases more sensitively with $\beta$. As $\beta$
approaches $2$, $p_{c}$ rapidly increases and takes on a value
close to 1, indicating that a rapidly aging network has the
slightly older nodes effectively isolated and hence only local
spread of a disease or rumor is possible, regardless of whether
links are further removed.  The phase boundary $p_{c}(\beta)$
separate the phase space into two regions, one corresponds to
local spread of a disease and another corresponds to a global
spread of a disease.  Note that for $\beta > \beta_{c}$, where
$\beta_{c} \approx 2.0$, $p_{c} =1$ implying that aging effect
{\em alone} is sufficient to keep a disease or rumor from
spreading globally.

The effect of aging can be understood qualitatively as follows. A
BA network without aging, i.e. $\beta=0$, is highly resilient to
random damages in the form of link removal \cite{albert2}.  The
results in Fig.1(a) show that one needs to remove more than $80\%$
of the links to prevent global spread of an epidemic.  This
feature is similar to previous studies of site percolation
\cite{cohen}, where it was shown that for networks with a
broad-scale degree distribution, one needs to remove a large
fraction of nodes (about $99\%$) before the network falls apart.
The sensitivity of $p_{c}$ to the aging parameter $\beta$ is
related to the form of the degree distribution as $\beta$
increases.  For small values of $\beta$ ($\beta < 1$), the degree
distribution still carries scale-free features that lead to a
small $p_{c}$ and the weak dependence of $p_{c}$ on $\beta$. For
larger values of $\beta$, the older nodes lose their
attractiveness rapidly giving rise to a narrow degree
distribution.  The scale-free nature of the network is lost and
the degree distribution becomes increasingly dominated by an
exponential form rather than a power law.  In this case, local
spread is more probable unless the network carries a larger
fraction of effective links, giving rise to the sensitive
dependence of $p_{c}$ on $\beta$ \cite{albert1,dorogovtsev}.  For
rapidly aging networks, the small world effect vanishes and the
``regular lattice" behavior dominates. The high clustering feature
of a rapidly aging network results in only local spread of a
diseases \cite{zheng}.

In summary, we have studied the combined effects of aging and
links removal on SIR dynamics in the Barab\'{a}si-Albert
scale-free networks. A threshold $p_{c}$ exists for the fraction
of effective links in a network below which only local spreads of
disease or rumor take place.  For small aging effect, $p_{c}$ is
low. As aging effect increases, $p_{c}$ increases.  For rapidly
aging network, $p_{c}=1$ implying that only local spread results,
regardless of the number of effective links in the network.

\acknowledgments  K.P.C. and P.M.H. would like to thank K. H.
Chung for useful discussions in the initial stage of this work.
One of us (D.F. Zheng) acknowledges the support from the National
Natural Science Foundation of China under grant number 70371069.

\newpage

\newpage
\centerline{\bf FIGURE CAPTIONS}

\bigskip
\noindent Fig. 1: The fraction of refractory nodes $r = N_{R}/N$
as a function of the size of network $N$ for different fractions
of effective links in BA networks on a log-log scale.  The aging
parameter is (a) $\beta = 0$, (b) $\beta = 0.5$, (c) $\beta = 1.0$
and (d) $\beta = 1.5$.  Note that for global spreads, $N_{R} \sim
N$ and a horizontal line results.

\bigskip
\noindent Fig. 2: The slope of $\log r$ against $\log N$ as a
function of fraction $p$ of effective links in a network. Note the
change from vanishing slope to negative slopes as $p$ decreases,
signifying a change from global to local spread of an epidemic.

\bigskip
\noindent Fig. 3: The thresholds $p_{c}(\beta)$ constitute a
boundary in the phase space formed by the aging parameter $\beta$
and fraction of effective links $p$.  In the region below (above)
the boundary, local (global) spreads occur.

\newpage
\begin{figure}
\epsfig{figure=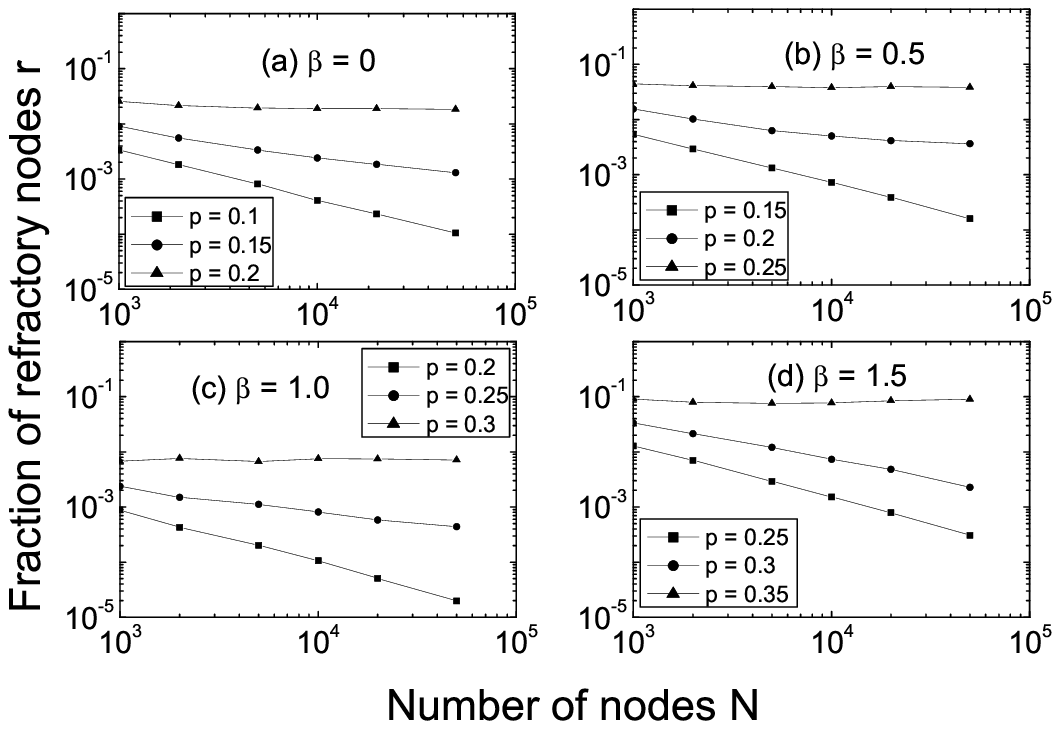,width=\linewidth} \label{figure1}
\end{figure}

\newpage
\begin{figure}
\epsfig{figure=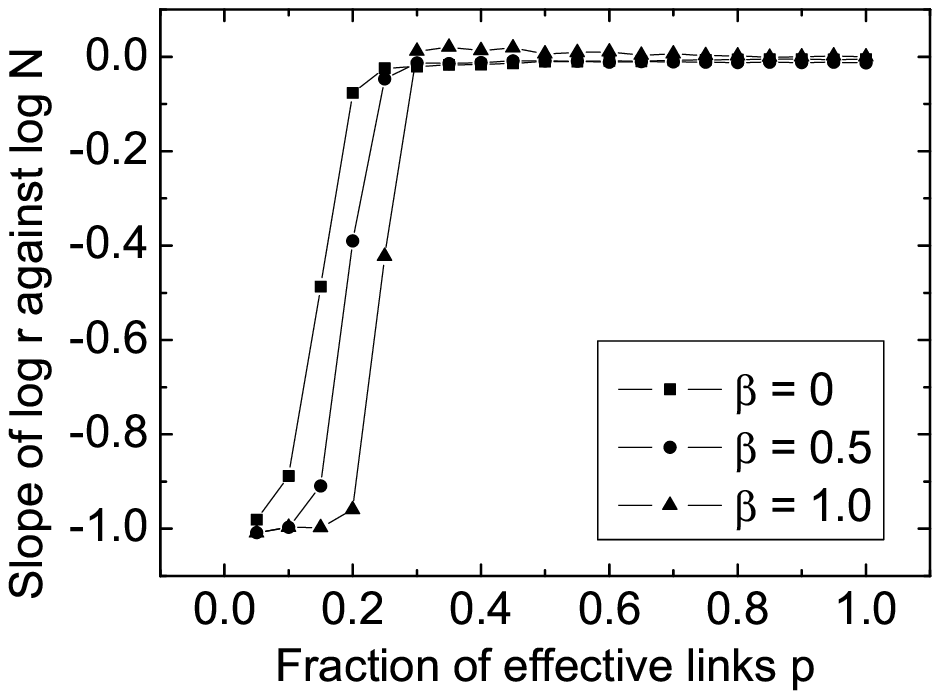,width=\linewidth} \label{figure2}
\end{figure}

\newpage
\begin{figure}
\epsfig{figure=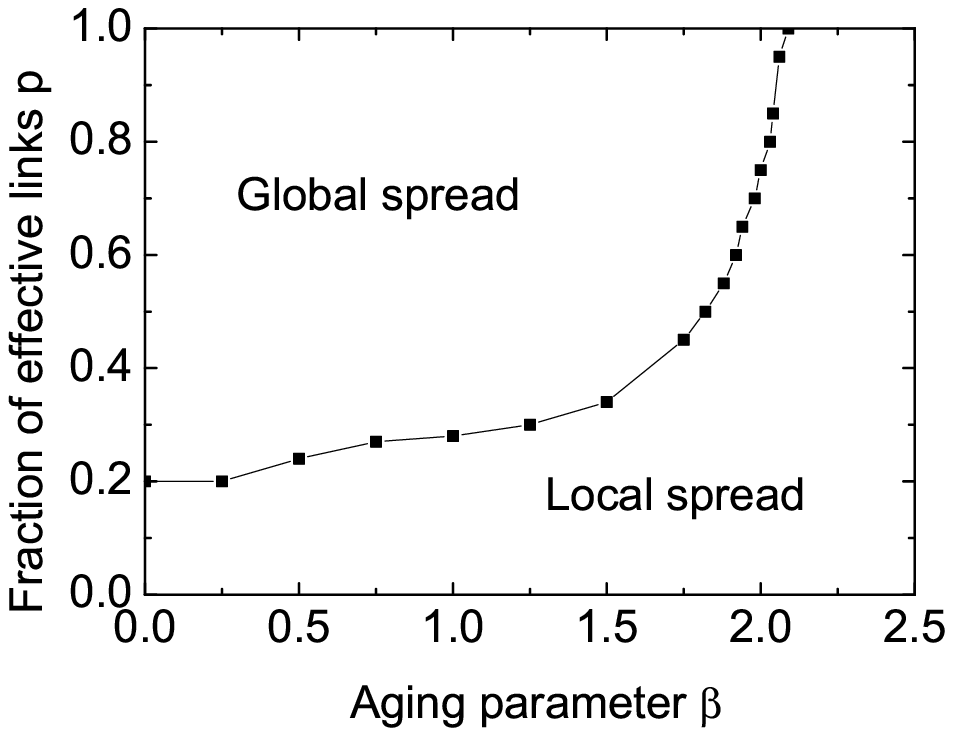,width=\linewidth} \label{figure3}
\end{figure}

\end{document}